\documentclass[11pt,equations,fleqn]{article}
\begin{document}
\begin{titlepage}
\def\baselinestretch{1.2}
\null
\begin{center}
{\LARGE
The Cluster Processor: New Results }
\vskip 10mm
{\Large
Henk W.J. Bl\"{o}te~$^{\S \dag}$\footnote{e-mail: bloete@tn.tudelft.nl},  
Lev N. Shchur~$^{\ddag}$\\  and Andrei L. Talapov~$^{\ddag}$
\footnote{present address:
Real3D, 2603 Discovery Drive, Suite 100, Orlando, FL 32825, USA},
\\}
\vskip 5mm
{\em
$^{\S}$ Faculty of Applied Science,  P.O. Box 5046,
2600 GA Delft, \\ The Netherlands \\
$^{\dag}$ Instituut Lorentz,
Universiteit Leiden, P.O. Box 9506, 2300 RA Leiden, The Netherlands\\
$^{\ddag}$ Landau Institute for Theoretical Physics,
142432 Chernogolovka, Russia\\}
\end{center}
\vskip 2mm
\def\baselinestretch{1.6}
\begin{abstract}
We present a progress report on the Cluster Processor, a special-purpose
computer system for the Wolff simulation of the three-dimensional Ising
model, including an analysis of simulation results obtained thus far.
These results allow, within narrow error margins, a determination of the
parameters describing the phase transition of the simple-cubic Ising
model and its universality class.  For an improved determination of
the correction-to-scaling exponent, we include Monte Carlo data for 
systems with nearest-neighbor and third-neighbor interactions in the
analysis.
\end{abstract}
\vfill
\end{titlepage}
\newcommand{\tm}{$\times$}

\section{Introduction}
The goal of the Cluster Processor project is to obtain an accurate
description of the phase transition of the three-dimensional Ising
model, in particular to determine the universal constants that apply
as well to many other phase transitions that occur in nature. The
Cluster Processor uses special-purpose hardware executing the Wolff
version \cite{W} of the `cluster' Monte Carlo algorithm first
introduced by Swendsen and Wang \cite{SW}.

The Cluster Processor project was planned to be
executed in two stages. The first stage was to construct, test, debug
and operate a single processor, called the Prototype Processor. The
second stage comprised the construction of a system of ten of such
processors that could be operated in parallel, which would thus allow
an even higher level of statistical accuracy.

The Monte Carlo results obtained thus far with help of the Cluster
Processor do indeed allow a numerical determination of the magnetic and
temperature renormalization exponents $y_h$ and $y_t$ with a level of
accuracy exceeding other Monte Carlo studies available to us at the time
that this paper was submitted. Nevertheless, the results are not yet
quite of the accuracy originally expected for the Cluster Processor.
Thus, we consider the Cluster Processor project as being yet unfinished.
Although the construction of hardware parts of this special-purpose
computer system, which includes a total of 11 processors, has already
been completed some time ago, only two of these were available in time
to contribute to the Monte Carlo analysis in the present article.
Two more processors have become operational in the meantime.

The construction part of the first stage was completed in 1995
\cite{TBS}, and subsequent simulations on the prototype processor 
yielded good results \cite{TBS,TB}.
The subsequent construction of the printed-circuit boards of 10
clones of the prototype processor was completed in about a year and a
half. However, during the last two years, the progress in the debugging
of the hardware has been rather slow. This problem is caused by the
absence of one of the one of us (A.L.T.) who built the hardware. This
author is now occupied by duties unrelated to the Cluster Processor.

In Section 2 we present a finite-size analysis of the Binder
cumulant \cite{Binder} of the 3-D Ising model
with nearest-neighbor interactions $K_{\rm nn}$ only. The data sample
consists of results taken with the Cluster Processor, mostly for finite
sizes $L=128$ and 256, supplemented by data for relatively small system
sizes, computed on workstations and personal computers. In Section 3
we extend the parameter space to include third-neighbor interactions
$K_{\rm 3n}$ which are known to reduce the corrections to scaling. The
numerical data for these systems are restricted to $L \leq 128$ and were
obtained for $K_{\rm 3n}/K_{\rm nn}=0.1$, 0.2, 0.3 and 0.4, using
workstations and personal computers. The Cluster Processor is
restricted to simulations of the nearest-neighbor model with finite
sizes $L=16$, 32, 64, 128 and 256. These additional data enable us
to perform an improved analysis, in particular concerning the accuracy
of the correction-to-scaling exponent. In Section 4 we present analyses
of the magnetic susceptibility and of the temperature-derivative of the
Binder cumulant \cite{Binder}, leading to new estimates of the
renormalization exponents $y_h$ and $y_t$. Section 5 discusses and
summarizes our findings; in particular see Table \ref{tab:sumres}.

The analysis presented in Section 3 is related to, and can be
seen as an extension of the analysis given in Ref. \cite{BLH} to a much
larger body of Monte Carlo data.
In the latter paper it was demonstrated that the amplitude of the
leading correction to scaling can be tuned by modifying the Hamiltonian:
in general, the use of a two-parameter Ising-like Hamiltonian enables one
to suppress both the temperature field and the leading irrelevant field.
In particular the latter condition can only be realized with a limited
numerical accuracy, so that it remains necessary to include the amplitude
of the leading
correction to scaling in the analysis. As emphasized in Ref. \cite{BLH},
the importance of reducing the irrelevant field lies in the fact that
the results of the fitting procedures then become almost independent of
the precise value of the irrelevant exponent $y_i$, which is known only
with a modest precision.

After the work of  Ref. \cite{BLH}, several new analyses of the 3-D Ising
model and its universality class have been performed. A modified Monte Carlo
renormalization analysis \cite{BHHMS}, which was set up such as to suppress
corrections to scaling, both at the simulated Hamiltonian and at the fixed
point of the transformation, yielded results for the critical exponents
with a precision comparable to those of  Ref. \cite{BLH}. An analysis of
the spontaneous magnetization and other data obtained by the prototype
Cluster processor led to somewhat better accuracies.
Very recently, two new Monte Carlo analyses have been reported. 
Ballesteros et al. \cite{Ballea} obtained very accurate numerical data
for the nearest-neighbor Ising model, and their results do indeed have
small error margins. Hasenbusch et al. \cite{Hasea} also
simulated the two-parameter Hamiltonians of Ref. \cite{BLH}, using a new
method to determine the parameters such as to suppress the leading
irrelevant field. They reported results with a precision comparable to
that of Ref. \cite{Ballea}. A further discussion of these new analyses,
and their similarities and dissimilarities with the present analysis will
be included in Section 4.

Further recent developments are the series expansion analysis of Butera
and Comi \cite{BCs} and by Campostrini et al. \cite{Campo}, and the
field-theoretic analysis of Guida and Zinn-Justin \cite{GZJ}. The latter
work presents the result $y_i=-0.81$ which is of some importance in
scaling analyses of Monte Carlo data.

\section{The nearest-neighbor Ising model}
We consider the Ising model with spins $s_{x,y,z}=\pm 1$ on the sites
$(x,y,z)$ of the simple-cubic lattice, with nearest-neighbor couplings
only:
\begin{equation}
  {\cal H}/kT = -K_{\rm nn} \sum_{x,y,z} s_{x,y,z} 
  (s_{x+1,y,z} +  s_{x,y+1,z} +  s_{x,y,z+1}) 
\end{equation}
The data used in this analysis of the nearest-neighbor Ising model
are the result of Cluster Processor simulations (roughly one half
processor-year), supplemented by some results for relatively small
system sizes, obtained with workstations and PC's. These simulations
used $L\times L\times L$ systems with periodic boundaries and took place
very close to the critical point, most of them at $K_{\rm nn}=0.2216545$.
The lengths of the Wolff simulations are shown in Table \ref{tab:simlen}.
This table presents the number of millions of samples \#S taken per system
size, and the number of Wolff steps \#W before taking each new sample.
One Wolff step means the formation and spin inversion of one cluster.
These data are shown only for $L\geq 40$ in Table \ref{tab:simlen}.
Also included in the present analysis are Monte Carlo data for some
smaller system sizes, namely $L=4$-16, 18, 20, 22, 24, 28 and 32.
These data, as well as the details of the simulations, were already 
described in Ref. \cite{BLH}.

\begin{table}[htbp]
\caption{Length of Monte Carlo runs for 5 different Ising models, in
millions of sampled configurations (\#S). The number \#W of Wolff
clusters flipped before a new configuration was sampled is also
indicated. Smaller system sizes $L<40$ are also included in the present
analysis (see text).
  }
\label{tab:simlen}
\begin{center}
\begin{tabular}{||r||r r|r r|r r|r r|r r||} 
\hline
$K_{\rm 3n}/K_{\rm nn}$
         & \multicolumn{2}{|c|}{0.0} &\multicolumn{2}{|c|}{0.1} &
                  \multicolumn{2}{|c|}{0.2} &\multicolumn{2}{|c|}{0.3} &
                                             \multicolumn{2}{|c|}{0.4} \\
\hline
$L$ &\#S & \#W   &\#S & \#W   &\#S & \#W   &\#S & \#W   &\#S & \#W   \\
\hline
 40 &100& 10 &100& 10 &100& 10 &100& 10 &100& 10 \\
 48 & 50& 20 &   &    & 50& 32 & 50& 32 &   &    \\
 64 & 50& 32 &   &    & 50& 32 & 50& 32 &   &    \\
128 & 18& 64 &   &    & 20& 64 & 20& 64 &   &    \\
128 &2.0&100 &   &    &   &    &   &    &   &    \\
256 &2.1&200 &   &    &   &    &   &    &   &    \\
\hline
\end{tabular}
\end{center}
\end{table}

The `data taking' during the simulations included the sampling of the
magnetization $m\equiv L^{-d} \sum_{x,y,x} s_{x,y,x}$ and the sum of
nearest-neighbor products (the energy of the nearest-neighbor model), and 
the accumulation of powers and cross-products of these. On this basis
we can obtain a number of expectation values. Theoretical expressions for
these quantities are also available. By matching these to the numerical
data, one can verify the theory and solve for unknown parameters such as
the location of the critical point.

The numerical data include the dimensionless ratio
\begin{equation}
  Q_L(K_{\rm nn})= \frac{\langle m^2\rangle_L^2}{\langle m^4\rangle_L}
\end{equation}
of moments of the magnetization $m$ per spin in a system of size $L$.
This ratio is a form of the Binder cumulant \cite{Binder}, a quantity that
has proven to be very useful in analyses of critical-point properties.
We copy the renormalization prediction for the scaling behavior of
$Q_L(K_{\rm nn})$ from Ref. \cite{BLH}:
%\pagebreak
\begin{displaymath}
  Q_L(K_{\rm nn})= Q + a_1 (K_{\rm nn}-K_{\rm c}) L^{y_{t}}+ a_2 (K_{\rm
    nn}-K_{\rm c})^2 L^{2y_{t}}+
\end{displaymath}
\begin{equation}
a_3 (K_{\rm nn}-K_{\rm c})^3 L^{3y_{t}} + \cdots+ \sum_j b_j L^{y_{j}} 
\label{expan}
\end{equation}
The sum on $j$ is expected to be dominated by a contribution $b_1 L^{y_1}$
where $y_1=y_i\approx -0.83$ \cite{NR,TB,GZJ} is the leading irrelevant
exponent. In addition, finite-size scaling predicts terms  \cite{BLH}
$b_2 L^{y_2}$ with  $y_2=3-2 y_h \approx -1.963$ and $b_3 L^{y_3}$ with
$y_3=y_t -2 y_h \approx -3.376$. The term with exponent $y_2$ is due to the
field-dependence of the analytic part of the free energy; a contribution
with exponent $y_3$ arises from nonlinear depencence of the
magnetic scaling field on the physical magnetic field. Furthermore, one
may expect higher order contributions from irrelevant fields; thus one
may also include a term  $q_1 b_1^2 L^{2 y_{i}}$.
Finite-size scaling also predicts further contributions to the sum in
Eq.~(\ref{expan}). Even if their exponents are smaller than $y_2$ or
$y_3$, this does not necessarily mean that they are less `important' than
$y_2$ or $y_3$, because the amplitudes may be very different. Which of these
contributions are incorporated in the `best fit' formula may be decided
on the basis of such requirements as that the residual $\chi^2$ of the 
least-squares analysis is acceptable, and that the fitted parameters
become reasonably independent of the minimum system size cutoff
$L_{\rm min}$ of the data included in the fit. A considerable number of
fits was tried out; in Table \ref{tab:fitdef} we 
define 8 different types by specifying which of the parameters were 
included in the minimization procedure. In those cases where the
irrelevant exponent $y_i$ was not fitted, it was set at the value
$y_i=-0.81$ taken from Guida and Zinn-Justin \cite{GZJ}.
\begin{table}[t]
\caption{Definition of several types of fits used in the analysis.
The parameters marked by `+' were actually solved in the minimization
process; the remaining ones were kept at a constant value (zero except
in the case of $y_i$, see text). }
\vskip 1 ex
\begin{center}
\begin{tabular}{||c|c|c|c|c|c|c|c||}
\hline
type & $Q$ & $K_c$  & $b_1$ &  $b_2$  & $b_3$ & $y_i$ & $q_1$ \\
\hline
  1  &  +  &   +    &   +   &         &       &       &       \\
  2  &  +  &   +    &   +   &         &       &   +   &       \\
  3  &  +  &   +    &   +   &    +    &       &       &       \\
  4  &  +  &   +    &   +   &    +    &       &   +   &       \\
  5  &  +  &   +    &   +   &    +    &   +   &       &       \\
  6  &  +  &   +    &   +   &    +    &   +   &   +   &       \\
  7  &  +  &   +    &   +   &    +    &       &   +   &   +   \\
  8  &  +  &   +    &   +   &    +    &   +   &   +   &   +   \\
\hline
\end{tabular}
\end{center}
\label{tab:fitdef}
\end{table}

The results of these fits are shown in Table \ref{tab:fitres1}.
For fits of type 1 and 2 it is necessary to discard an appreciable number
of the smaller system sizes. It appears to be difficult to estimate the
universal constant $Q$ accurately; its variation between different
fits exceeds some of the statistical error estimates.

\begin{table}[t]
\caption{ Results of least-squares fits to the
dimensionless amplitude ratio $Q$, using the simulation data of the 
nearest-neighbor Ising model. Different types of fits are used;
the type of fit is indicated in column 1. Data for system sizes below
the  specified minimum size (column 2) were ignored. Columns 3-5
show the results for the critical point, the Binder cumulant and the
irrelevant exponent. Where indicated by `(f)', $y_i$ was fixed at a
constant value $-0.81$. The last two columns show the residuals and
the number $d_f$ of degrees of freedom of the fits respectively.
}
\vskip 1 ex
\begin{center}
\begin{tabular}{||r|r|lr|lr|lr|l|l||}
\hline
type&$L\geq$& $K_c$      &     &   $Q$   &     & $y_i$ &   &$\chi^2$& $d_f$\\
\hline
  1 &   12  & 0.22165434 & (5) & 0.62261 & (5) & -0.81 &(f)& 54.4   & 38   \\
  1 &   14  & 0.22165438 & (5) & 0.62270 & (7) & -0.81 &(f)& 45.6   & 36   \\
  1 &   16  & 0.22165442 & (5) & 0.62280 & (8) & -0.81 &(f)& 38.8   & 33   \\
  2 &   10  & 0.22165462 & (6) & 0.6240  & (2) & -0.95 &(2)& 43.1   & 38   \\
  2 &   12  & 0.22165456 & (7) & 0.6237  & (2) & -0.91 &(2)& 35.7   & 34   \\
  2 &   14  & 0.22165457 & (8) & 0.6237  & (3) & -0.92 &(3)& 35.2   & 32   \\
  3 &    8  & 0.22165451 & (6) & 0.62320 & (8) & -0.81 &(f)& 50.3   & 51   \\
  4 &    7  & 0.22165459 & (7) & 0.6238  & (3) & -0.91 &(4)& 55.0   & 52   \\
  4 &    8  & 0.22165453 & (8) & 0.6233  & (4) & -0.83 &(6)& 50.2   & 50   \\
  5 &    4  & 0.22165455 & (5) & 0.62338 & (7) & -0.81 &(f)& 56.1   & 56   \\
  5 &    5  & 0.22165457 & (6) & 0.62344 & (9) & -0.81 &(f)& 54.7   & 55   \\
  5 &    6  & 0.22165456 & (6) & 0.62341 &(11) & -0.81 &(f)& 54.5   & 54   \\
  6 &    4  & 0.22165456 & (8) & 0.6234  & (3) & -0.82 &(5)& 56.0   & 55   \\
  6 &    5  & 0.22165450 & (9) & 0.6230  & (5) & -0.74 &(7)& 53.7   & 54   \\
\hline
\end{tabular}
\end{center}
\label{tab:fitres1}
\end{table}

\section{Including third-neighbor interactions}
We use the same reduced spin-one-half Hamiltonian as in Ref. \cite{BLH},
namely
\begin{displaymath}
  {\cal H}/kT = -\sum_{x,y,z} s_{x,y,z} \{
K_{\rm nn}(s_{x+1,y,z} +  s_{x,y+1,z} +  s_{x,y,z+1}) +
\end{displaymath}
\begin{equation}
K_{\rm 3n}(s_{x+1,y+1,z+1}+s_{x+1,y-1,z-1}+s_{x-1,y+1,z-1}+s_{x-1,y-1,z+1})\}
\end{equation}
Whereas the analysis of Ref. \cite{BLH} was restricted to only one
nonzero value $K_{\rm 3n}/K_{\rm nn}=0.4$, here we include similar Monte
Carlo data for $K_{\rm 3n}/K_{\rm nn}=0.1$, 0.2 and 0.3 as well. For the
smaller system sizes, the run lengths and further details of the
simulations are mostly equal to those specified in Ref. \cite{BLH} for
$K_{\rm 3n}/K_{\rm nn}=0.4$. For system sizes $L>32$, the lengths of
the Monte Carlo runs are included in Table \ref{tab:simlen}. The focus   
of these additional simulations was placed on the cases
$K_{\rm 3n}/K_{\rm nn}=0.2$ and 0.3, where the leading correction to
scaling becomes small.

Some fits of the types defined in Table \ref{tab:fitdef} were applied 
to the $Q$ data. The results were found to agree satisfactorily with
universality of $Q$, so that, following Ref. \cite{BLH}, the combined $Q$
data for the 5 models were analyzed by a single fit formula Eq.~(\ref{expan})
in which the universal parameters appear only once, and the nonuniversal
ones in 5-fold. This was done for several choices of combinations of
parameters as defined in Table \ref{tab:fitdef}, in particular those types
that include the irrelevant exponent $y_i$ as a free parameter. Some of
the results are shown in Table \ref{tab:fitres2}.

\begin{table}[t]
\caption{ Results of least-squares fits for the dimensionless amplitude
ratio $Q$, using the combined simulation data for 5 Ising models.
Different types of fits are used (see text); the type is indicated in
column 1. Data for system sizes below the specified  minimum size
(column 2) were ignored. The results for $K_c$ apply to the
nearest-neighbor model.
}
\vskip 1 ex
\begin{center}
\begin{tabular}{||r|r|lr|lr|lr|l|l||}
\hline
type&$L\geq$& $K_c$      &     &   $Q$   &     & $y_i$ &   &$\chi^2$& $d_f$\\
\hline
  2 &   14  & 0.22165431 & (5) & 0.62234 & (8) &-0.780& (9)& 148.2  &  64  \\
  2 &   16  & 0.22165437 & (5) & 0.62249 &(10) &-0.781&(12)& 105.7  &  53  \\
  2 &   18  & 0.22165441 & (5) & 0.62267 &(12) &-0.787&(16)&  79.0  &  47  \\
  4 &    5  & 0.22165444 & (5) & 0.62342 & (4) &-0.906& (6)& 361.2  & 164  \\
  4 &    6  & 0.22165452 & (5) & 0.62352 & (5) &-0.887& (8)& 221.4  & 159  \\
  4 &    7  & 0.22165454 & (5) & 0.62348 & (6) &-0.859&(11)& 186.4  & 153  \\
  4 &    8  & 0.22165457 & (5) & 0.62348 & (7) &-0.839&(15)& 175.6  & 147  \\
  6 &    4  & 0.22165462 & (5) & 0.62370 & (6) &-0.849&(12)& 193.8  & 164  \\
  6 &    5  & 0.22165460 & (5) & 0.62358 & (8) &-0.824&(17)& 184.8  & 159  \\
  6 &    6  & 0.22165458 & (5) & 0.62341 &(11) &-0.799&(24)& 177.2  & 154  \\
  7 &    5  & 0.22165455 & (5) & 0.62367 & (5) &-0.815& (7)& 213.3  & 163  \\
  7 &    6  & 0.22165459 & (5) & 0.62367 & (6) &-0.814&(12)& 186.3  & 158  \\
  7 &    7  & 0.22165459 & (5) & 0.62357 & (7) &-0.797&(19)& 176.0  & 152  \\
  8 &    4  & 0.22165463 & (5) & 0.62372 & (7) &-0.834&(22)& 193.2  & 163  \\
  8 &    5  & 0.22165460 & (5) & 0.62358 & (9) &-0.826&(37)& 184.5  & 158  \\
  8 &    6  & 0.22165459 & (6) & 0.62344 &(12) &-0.760&(56)& 176.7  & 153  \\
\hline
\end{tabular}
\end{center}
\label{tab:fitres2}
\end{table}
The simultaneous analysis of these 5 models is seen to allow a somewhat
more precise determination of $Q$ and $y_i$.
Again, the fit of type 2, involving only one correction to scaling, is
seen to converge too slowly as a function of the minimum system size.
The convergence is seen to improve when the number of parameters 
increases. However, the fits of types 7 and 8, which include a term
$q_1 b_1^2 L^{2 y_{i}}$ (not 5, only one because $q_1$ is universal)
in addition to those already given in Eq.~\ref{expan}, did not provide
strong evidence for such a nonlinear contribution in the leading
irrelevant scaling field.

\section{The magnetic and the temperature exponents}
A factor $kT$ is included in the susceptibility, so that the
high-temperature dependence vanishes and $\chi=1$ according to Curie's law.
It is thus expressed in magnetization fluctuations as
\begin{equation}
 \chi = L^d  \langle m^2 \rangle \;.
\end{equation}
The expected finite-size scaling behavior is
\begin{displaymath}
  \chi = c_0 + c_1 (K_{\rm nn}-K_{\rm c}) + \cdots +
L^{2y_{\rm h}-d} \left[ a_0 +
\right.
\end{displaymath}
\begin{equation}
\left.
a_1(K_{\rm nn}-K_{\rm c})L^{y_{\rm t}} + a_2(K_{\rm nn}-K_{\rm c})
L^{2y_{\rm t}} + b_1 L^{y_{\rm i}} + b_2 L^{y_{\rm j}} \cdots \right] \;
\label{eq:chiscal}
\end{equation}
Several types of fit were applied to the numerical susceptibility data for
the 3-d nearest-neighbor Ising model. In all fits, the critical point was
fixed at $K_{\rm c}=0.22165459$ and the irrelevant exponent at $y_i=-0.81$.
In fits of type A (see Table \ref{tab:fitres3}) we set $b_2=0$ in
Eq.~(\ref{eq:chiscal}); in types 2 and 3, it was left free. The associated
second correction exponent was fixed at $y_j= y_t -2y_h$ (type B) or at
$y_j= -2y_h$ (type C). The type B value follows when one includes a
quadratic term in the physical magnetic field in the temperature-like
scaling field; the type C value represents the field dependence of
finite-size contribution to the critical free-energy density.

\begin{table}[t]
\caption{ Results of least-squares fits to the finite-size data for the
susceptibility of the nearest-neighbor Ising model.
The type of fit is indicated in column 1. The minimum system sizes used
in the fits are shown in column 2.
}
\vskip 1 ex
\begin{center}
\begin{tabular}{||r|r|lr|c|c||}
\hline
type&$L\geq$& $y_h$      &      & $\chi^2$& $d_f$\\
\hline
  A &    5  & 2.48222    & (17) &  52.0   & 37   \\
  A &    6  & 2.48182    & (21) &  31.1   & 36   \\
  A &    7  & 2.48164    & (25) &  27.6   & 35   \\
  A &    8  & 2.48149    & (29) &  25.8   & 34   \\
  B &    4  & 2.48151    & (25) &  30.0   & 37   \\
  B &    5  & 2.48131    & (31) &  27.7   & 36   \\
  B &    6  & 2.48139    & (38) &  27.4   & 35   \\
  C &    4  & 2.48179    & (21) &  33.9   & 37   \\
  C &    5  & 2.48148    & (27) &  26.9   & 36   \\
  C &    6  & 2.48145    & (32) &  26.9   & 35   \\
\hline
\end{tabular}
\end{center}
\label{tab:fitres3}
\end{table}

As explained in Ref. \cite{BLH} it is possible to obtain the temperature
derivative of the Binder cumulant by sampling magnetization moments, the
energy  and their correlations. This quantity is expected to scale as
\begin{displaymath}
  \frac{\partial Q}{\partial K_{\rm nn}} = L^{y_{\rm t}} \left[ u_0+u_1
  (K_{\rm nn}-K_{\rm c})L^{y_{\rm t}}+u_2(K_{\rm nn}-K_{\rm c})^2
  L^{2y_{\rm t}}+ \cdots + \right.
\end{displaymath}
\begin{equation}
  \left.
  v_1 L^{y_{\rm i}} + v_2 L^{y_j} + \cdots
  \right]\;.
\label{qdevsc}
\end{equation}

Several types of fit were applied to the Monte Carlo data obtained for
this quantity in the case of the 3-d nearest-neighbor Ising model. The
critical point was fixed at $K_{\rm c}=0.22165459$ and the irrelevant
exponent at $y_i=-0.81$. In fits of type a (see Table \ref{tab:fitres4})
we set $v_2=0$ in
Eq.~(\ref{qdevsc}); in types b, c and d it was left free. The associated
second correction exponent was fixed at $y_j=3-2 y_h $ (type b) or at
$y_j=y_t -2 y_h$ (type c). The type b value is due to the field-dependence
of the analytic part of the free energy, and the type c value follows
when one includes a quadratic term in the physical magnetic field in
the temperature-like scaling field. Both terms were included in fits of
type d.

\begin{table}[t]
\caption{ Results of least-squares fits to the finite-size data for the
temperature derivative of the Binder cumulant of the nearest-neighbor
Ising model.
The type of fit is indicated in column 1. The minimum system sizes used
in the fits are shown in column 2.
}
\vskip 1 ex
\begin{center}
\begin{tabular}{||r|r|lr|c|c||}
\hline
type&$L\geq$& $y_h$      &      & $\chi^2$& $d_f$\\
\hline
  a &    5  & 1.5841     &  (3) &  50.5   & 43   \\
  a &    6  & 1.5845     &  (4) &  45.9   & 42   \\
  a &    7  & 1.5847     &  (4) &  45.4   & 41   \\
  a &    8  & 1.5852     &  (5) &  41.2   & 40   \\
  a &    9  & 1.5857     &  (5) &  37.5   & 39   \\
  a &   10  & 1.5858     &  (6) &  37.1   & 38   \\
  b &    4  & 1.5860     &  (5) &  40.7   & 43   \\
  b &    5  & 1.5861     &  (6) &  40.6   & 42   \\
  b &    6  & 1.5866     &  (7) &  40.2   & 41   \\
  b &    7  & 1.5872     &  (9) &  37.4   & 40   \\
  c &    4  & 1.5852     &  (4) &  43.1   & 43   \\
  c &    5  & 1.5854     &  (5) &  42.3   & 42   \\
  c &    6  & 1.5859     &  (6) &  41.3   & 41   \\
  c &    7  & 1.5865     &  (7) &  36.8   & 40   \\
  c &    8  & 1.5867     &  (8) &  36.5   & 39   \\
  d &    4  & 1.5865     &  (9) &  40.2   & 42   \\
\hline
\end{tabular}
\end{center}
\label{tab:fitres4}
\end{table}
The fits of type a display a slow trend as a function of $L$ so that
additional corrections should be taken into account. The residuals
do not clearly discriminate between type b or c corrections. The results
for both types agree well with that for type d.

\section{Discussion}
A major factor limiting the accuracy of Monte Carlo analyses of critical
phenomena is the multitude of correction-to-scaling mechanisms that 
should be included in a proper theoretical description, i.e. in the
fit formula that is used to model the finite-size data.
Each correction that is neglected
in the data analysis will lead to a bias in the results of the analysis.
Unfortunately, the statistical accuracy of the estimated parameters tends
to decrease rapidly when more free parameters are introduced. It is thus
necessary to restrict the number of free parameters. One may  estimate
the effect of the neglected corrections by imposing a minimim system size
$L_{\rm min}$, ignoring data for $L<L_{\rm min}$, and observing the way 
in which the residual $\chi^2$ depends on $L_{\rm min}$. Fortunately,
$\chi^2$ tends to decrease rapidly when $L_{\rm min}$ increases, especially
when the number of parameters is not too small.

Very few free parameters appear in the fits of types 1 and 2 in
Tables \ref{tab:fitres1} and \ref{tab:fitres2}: only one finite-size
correction is taken into acount. This is
reflected by the large values of $L_{\rm min}$ needed for 
acceptable fits, and by the large differences of e.g. the irrelevant
exponent $y_i$ with respect to the expected value $\approx -0.82$.
Similar effects are seen in the recent analysis of Ballesteros
et al. \cite{Ballea} who also use one correction. The effects on our
present analysis are more serious  because the data have a somewhat
higher statistical accuracy.

Thus, we found it necessary to include a second correction in the analysis
of the Binder cumulant, the term with exponent $y_2=3-2 y_h$ in
Eq.~(\ref{expan}). This leads to a large reduction of the residual
$\chi^2$; the fits (of types 3 and 4) are now seen to
become acceptable already at $L_{\rm min}=8$ and then yield reasonable
values for $y_i$.

It is remarkable that the term with exponent $y_2$ was not present in
the analysis given by Hasenbusch et al. \cite{Hasea}, which nevertheless
included small system sizes. This term is likely to be responsible for
the `slow convergence of the NNN flow' observed by Hasenbusch et al. in
their analysis of an Ising model with first and third neighbor interactions.
No such ill effect was however observed when the same analysis was applied
to a spin-one model \cite{Hasea}.
The different behavior of these two models is in agreement with results
already given in Ref. \cite{BLH} which pertain to models that do not
differ much from those of Ref. \cite{Hasea}. These results include the
finite-size amplitudes of the $L^{y_2}$ term ($b_2$ in Tables 5 and 6 of
Ref. \cite{BLH}); this amplitude is appreciable for the Ising model with
first and third neighbor interactions, and approximately zero for the
spin-one model.

The present results in Tables \ref{tab:fitres1} and \ref{tab:fitres2}
 illustrate that a further
strong reduction of $\chi^2$ occurs when a third correction term is
included. The fits of types 5, 6 and 8 are able to describe the 
finite-size dependence of $Q$ for even smaller values of $L$.
It is not completely clear that the term with exponent $y_3= y_t-2y_h$
is actually the most important that has to be accounted for. There
may also be significant corrections due to a second irrelevant field
with an unknown exponent, and due to the expected nonlinear dependence
of the magnetic scaling field on the physical field. The latter
mechanism produces a finite-size correction in $Q$ with exponent
$y_4=-2y_h$ which is even smaller than $y_3$. Fortunately the important
fitted parameters do not depend strongly on the choice between $y_3$
and $y_4$. This observation gives us some more confidence that the
statistical error estimates are reasonable indicators of the actual
uncertainties. We still prefer to give conservative two-sigma error 
bounds, in order to allow for the arbitrariness in the form of the
fit formulas. Our preferred values and error estimates are summarized
in Table \ref{tab:sumres}.

\begin{table}[t]
\caption{ Final results of the present work. The argument of the
quantities given in the first column is the coupling ratio
$K_{\rm 3n}/K_{\rm nn}$. The values of the critical points $K_c$
apply to the nearest-neighbor coupling $K_{\rm nn}$. The amplitude
$b_1$ of the leading finite-size correction in $Q$ was determined 
assuming a fixed irrelevant exponent $y_i=-0.81$.
Two-sigma error estimates are given in the rightmost column.
}
\vskip 1 ex
\begin{center}
\begin{tabular}{||l|l r||}
\hline
quantity      &     value  &      error \\
\hline
$K_c(0.0)$  & 0.22165459 & (10)  \\
$K_c(0.1)$  & 0.18562466 & (52)  \\
$K_c(0.2)$  & 0.16073235 & (12)  \\
$K_c(0.3)$  & 0.14230187 & (12)  \\
$K_c(0.4)$  & 0.12800393 & (40)  \\
$b_1(0.0)$  & 0.0896     & (25)  \\
$b_1(0.1)$  & 0.0494     & (25)  \\
$b_1(0.2)$  & 0.0142     & (25)  \\
$b_1(0.3)$  &-0.0162     & (25)  \\
$b_1(0.4)$  &-0.0465     & (25)  \\
$Q$         & 0.62358    & (15)  \\
$y_h$       & 2.4814     &  (5)  \\
$y_t$       & 1.5865     & (14)  \\
$y_i$       & -0.82      &  (3)  \\
\hline
\end{tabular}
\end{center}
\label{tab:sumres}
\end{table}

The result for the critical point of the nearest-neighbor model 
agrees well with that of Ballesteros et al. \cite{Ballea} and
reasonably with that of Hasenbusch et al. \cite{Hasea}. Also the
universal quantities given in Table \ref{tab:sumres} are
mostly consistent with Refs. \cite{Ballea} and \cite{Hasea} but we
note a rather large difference with $Q$ of Ref. \cite{Hasea}. A 
possible explanation of this difference is the presence of the term
$b_2 L^{y_{2}}$ in Eq.~(\ref{expan}) which was not taken into account
in Ref. \cite{Hasea}. Although the amplitude $b_2$ is small \cite{BLH},
we see no reason why it should precisely be zero.

Although our present work involves a body of simulations that is 
larger than those of Refs. \cite{Ballea} and \cite{Hasea}, this is not
reflected in a smaller error bar for $y_h$ (see Table \ref{tab:sumres})
than those given in Refs. \cite{Ballea} and \cite{Hasea}. The main
reason is obviously that our quoted errors are two standard deviations.
The errors given in Refs. \cite{Ballea} and \cite{Hasea} should
be multiplied by two before comparing with our work.

Furthermore we note that our results for the critical exponents agree well
with recent series-expansion results of Campostrini et al. \cite{Campo}.
Finally, our results for $y_i$ and $y_t$ are in a good agreement
with the field-theory results obtained recently by Guida and
Zinn-Justin \cite{GZJ}.  Although the agreement is not so good in the
case of $y_h$ where the difference is slightly larger than the
combined error bars, the overall agreement between the results of such
different approaches is quite satisfactory and confirms the underlying
assumptions of scaling and universality.

{\em Acknowledgements}: 
It is a pleasure to thank M.E. Fisher, J.R. Heringa, J.M.J. van Leeuwen,
E. Luijten and W. Selke for valuable discussions.
This research is supported by the NWO ('Nederlandse Organisatie voor
Wetenschappelijk Onderzoek') via grants \# 047-13-210 and \# 047-003-019,
by INTAS via grant \# 93-211, by 
the FOM ('Stichting voor Fundamenteel Onderzoek der Materie') which is
financially supported by the NWO, and partially by RFBR (Russian
Foundation for Basic Research) via grant \# 99-02-18312.

\end{document}